# Finite-size scaling at a topological transition: Bilinear-biquadratic spin-1 chain

Yuting Wang,[1] Hao Zhang,[1] and Alex Kamenev[1,2]
[1]*School of Physics and Astronomy, University of Minnesota, Minneapolis, Minnesota 55455, USA*
[2]*William I. Fine Theoretical Physics Institute, University of Minnesota, Minneapolis, Minnesota 55455, USA*

We consider a finite-size scaling function across a topological phase transition in one-dimensional models. For models of noninteracting fermions it was shown to be universal for all topological symmetry classes and markedly asymmetric between trivial and topological sides of the transition [T. Gulden, M. Janas, Y. Wang, and A. Kamenev, Phys. Rev. Lett. **116**, 026402 (2016)]. Here we verify its universality for the topological transition between dimerized and Haldane phases in bilinear-biquadratic spin-1 chain. To this end we perform high-accuracy variational matrix product state simulations. We show that the scaling function, expressed in terms of $L/\xi$, where $L$ is the chain length and $\xi$ is the correlation length, coincides with that of three species of noninteracting massive Majorana fermions. The latter is known to be a proper description of the conformal critical theory with central charge $c = 3/2$. We have shown that it still holds away from the conformal point, including the finite-size corrections. We have also observed peculiar differences between even- and odd-size chains, which may be fully accounted for by residual interactions of the edge states.

## I. INTRODUCTION

Topological states of matter continue to attract ever increasing attention of the community [1–4], vis-a-vis their peculiar electric and thermal transport properties as well as applications in quantum computing. Yet, surprisingly little attention was paid to the most basic thermodynamical quantities and their scaling properties close to topological phase transitions. Though hard to measure, they exhibit a remarkable universality and provide a conceptual framework to distinguish between different universality classes.

In this paper we discuss a finite-size scaling of a many-body ground-state energy across topological phase transitions in $1 + 1$ dimensions. Critical points of such models are described by conformal field theories (CFT) [5], characterized by the central charge $c$. The finite-size, $L$, scaling of the ground-state energy $E(L, \infty)$ for an *open* system at criticality was shown to be [6,7]

$$E(L, \infty) = L\bar{\epsilon}(\infty) + b(\infty) - \frac{c}{L}\frac{\pi}{24} + O(L^{-2}), \quad (1)$$

where $\bar{\epsilon}(\infty)$ is the average bulk energy density, $b(\infty)$ is a size-independent boundary term, and the argument $(\infty)$ specifies the exact critical point where the correlation length $\xi \to \infty$. Here the velocity of excitations ("Fermi" velocity) is put to be 1. The $1/L$ term appears to be universal and depends only on $c$—the central charge of the Virasoro algebra.

A relevant perturbation drives the system away from criticality, creating a spectral gap $\Delta$ and a corresponding correlation length $\xi = 1/\Delta$. One may generalize the CFT expansion, Eq. (1), as

$$E(L, \xi) = L\bar{\epsilon}(\xi) + b(\xi) - \frac{c}{L}f\left(\frac{L}{\xi}\right) + O(L^{-2}). \quad (2)$$

The first two terms on the right side are well defined for any fixed $\Delta$ or $\xi$ by studying the asymptotic limit $L \gg \xi$ [we will see that in this limit $f(L/\xi)$ is exponentially small]. Once $\bar{\epsilon}(\xi)$ and $b(\xi)$ are known one may study the double scaling limit [8]: $L \to \infty$ and $\xi \to \infty$, while $w = L/\xi = $ const. The scaling function $f(w)$ is then defined as

$$cf(w) = \lim_{w\xi=L\to\infty} L(L\bar{\epsilon}(\xi) + b(\xi) - E(L, \xi)). \quad (3)$$

According to Eq. (1), $f(0) = \pi/24$. Universality of the scaling function $f(w)$ for $w \neq 0$ and its ability to distinguish between topological sectors is the subject of this work.

The scaling function was studied [9] for the class of $1 + 1$-dimensional topological models of noninteracting fermions. It was shown that it is universal for all symmetry classes, admitting nontrivial topology in one dimension [10–12]: AIII, DIII, and CII, where $c = 1$; and BDI and D, where $c = 1/2$ (see Fig. 1). Moreover, it was shown that the corresponding $f(w)$ may be derived from the Dirac Hamiltonian, e.g., in the AIII symmetry class ($c = 1$), $\mathcal{H} = m\sigma_1 + i\partial_x\sigma_2$, where the Pauli matrices act in the sublattice $A/B$ space. The model is equivalent to *two* copies of $c = 1/2$ Majorana fermions. Assuming that outside of the interval $0 < x < L$ the gap is very large and, e.g., negative, one derives the boundary conditions $\Psi_A(0) = \Psi_B(L) = 0$. The quantized values of momenta $k > 0$ are then given by

$$\cos[kL + \delta(k)] = 0, \quad \tan\delta(k) = \frac{m}{k} = \frac{w}{kL}. \quad (4)$$

As a result the spectrum is determined by the condition $w \equiv Lm = k_nL\cot(k_nL)$ and the energies are given by $\epsilon^\pm(k_n) = \pm\sqrt{m^2 + k_n^2}$. At $w = 1$ two of its real solutions collide and switch to purely imaginary ones for $w > 1$. Those correspond to the topological edge states, decaying into the bulk of the system.

The total ground-state energy is given by $E(L, \xi) = \sum_n \epsilon^-(k_n)$, which, using the argument principle, may be

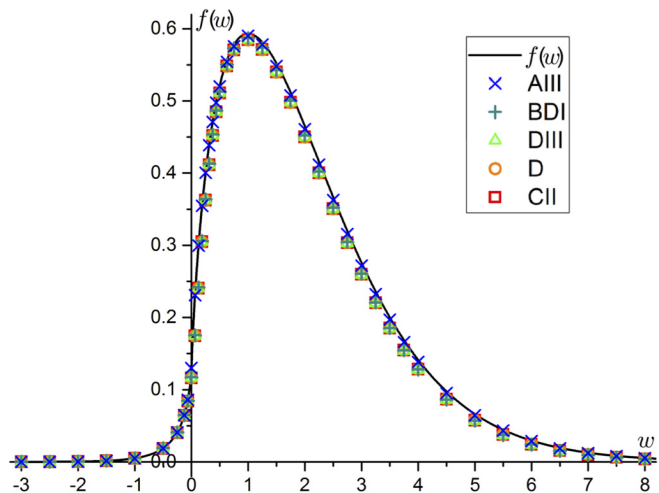

FIG. 1. The solid line is the scaling function equation, Eq. (6). Here $w > 0$ ($<0$) represents the topological nontrivial (trivial) side of the transition. (The edge states exist for $w > 1$, see the main text.) Symbols are numerical results for lattice models of noninteracting fermions in five symmetry classes; after Ref. [9].

written as

$$E(L, \xi) = \frac{1}{2} \oint \frac{dk}{2\pi i} \epsilon^-(k) \, \partial_k \ln \{\cos[kL + \delta(k)]\}, \quad (5)$$

where the contour runs in the complex $k$-plane encircling all solutions of Eq. (4). The bulk and boundary terms are given by $L\bar{\epsilon} + b = \int (dk/2\pi)\epsilon^-(k)[L + \partial_k\delta(k)]$, where $L + \partial_k\delta(k)$ are bulk and boundary parts of the continuous density of states. To find the scaling function $f(w)$, one employs Eq. (3), deforms the integration contour to run along the branch cut of $\sqrt{m^2 + k^2}$, and rescales the integration variable as $z = ikL$. As a result, one finds [9]

$$f(w) = -\int_{|w|}^{\infty} \frac{dz}{\pi} \sqrt{z^2 - w^2} \, \partial_z \ln[1 + e^{-2z - 2\delta_w(z)}], \quad (6)$$

where $\delta_w(z) = -\text{arctanh}(w/z)$. This expression is plotted as a solid line in Fig. 1.

One may see that the scaling function is markedly asymmetric between the topological nontrivial, $w > 0$, and the topological trivial, $w < 0$, sides. This is a feature of an open system. Indeed, similar calculation for periodic boundary conditions results in a symmetric function [9]. On the other hand, a specific shape of the boundary (e.g., a shape of the gap $m(x)$ near the boundary) does not change the scaling function. This suggests that the asymmetry is due to the presence of the edge states on the topological nontrivial side of the transition. This is corroborated with the fact that the maximum of the scaling function (i.e., maximum sensitivity to the finite-size effects) occurs at $w = 1$, i.e., $\xi = L$, which is exactly the point where the edge states (wave function $e^{ik_n x}$ with purely imaginary wave number $k_n$) appear. Glancing at the scaling function, one would be hard-pressed to locate the point of the bulk-phase transition. However, a more accurate look reveals a nonanalytic behavior of the form $f(w) \approx \frac{\pi}{24} - \frac{w}{\pi} \log |w|$ close to $w = 0$, marking the bulk transition point.

Our goal here is to verify if the scaling function equation, Eq. (6), is applicable beyond the simple models of noninteracting fermions [9]. To this end we evaluate it for bilinear-biquadratic spin-1 chain, using the variational matrix product state (MPS) approach [13–15] and the variational uniform MPS (VUMPS) algorithm [16]. Unless explicitly stated otherwise, we study the open chain with an even number of sites case. The model undergoes a topological phase transition between dimerized and Haldane phases [17]. The transition is known [18] to be described by a CFT with $c = 3/2$.

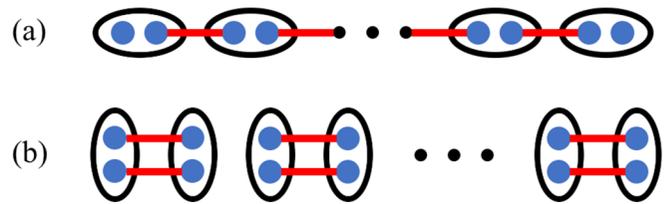

FIG. 2. Schematic representation of (a) the Haldane phase and (b) the dimerized phase. Each spin-1 is represented as two spin-1/2 objects, red links indicate singlet bonds.

We conclude that, to the best of our numerical precision, the scaling function of bilinear-biquadratic spin-1 chain is indeed in agreement with the analytical result, Eq. (6). This fully supports the theory [18–22] that the low-energy physics of the model is equivalent to that of three species of massive Majorana fermions. According to our results, this correspondence goes beyond the bulk of the spectrum and encompasses the finite-size physics, including the edge states. It is probably exact in the double scaling limit, Eq. (3). Though the scaling function, expressed through $L/\xi$, coincides with the free fermion one, the correlation length $\xi$ exhibits a rather nontrivial dependence on parameters of the model (due to the presence of marginal operators).

The paper is organized as follows: In Sec. II we introduce the model and discuss our numerical results for the scaling function in the even-sites case. The odd-size chains are discussed in Sec. III. In Sec. IV we describe the variational MPS approach and the VUMPS algorithm as well as details of our numerical approach. Finally conclusions and open questions are summarized in Sec. V.

## II. THE MODEL AND SCALING FUNCTION

The model we study is the bilinear-biquadratic spin-1 chain with the Hamiltonian

$$\mathcal{H} = \sum_{i=1}^{L} \cos\theta (\vec{S}_i \cdot \vec{S}_{i+1}) + \sin\theta (\vec{S}_i \cdot \vec{S}_{i+1})^2, \quad (7)$$

where $\vec{S}_i$ is the spin-1 operator at site $i$ and $\theta$ is a parameter that controls the relative strength between the bilinear and biquadratic interactions. This model exhibits a rich phase diagram when $\theta$ is varied between $-\pi$ and $\pi$. In particular, $\theta = 0$ is the Heisenberg point [13], and $\theta = \arctan\frac{1}{3}$ is the Affleck-Kennedy-Lieb-Tasaki model [23] with a known exact ground state.

The quantum phase transition we concentrate on is between the Haldane phase, $-\frac{\pi}{4} < \theta < \frac{\pi}{4}$, and the dimerized phase, $-\frac{3\pi}{4} < \theta < -\frac{\pi}{4}$ (see Fig. 2). The system is gapped in the Haldane phase with a unique ground state under the periodic boundary condition (PBC) and fourfold degenerate (in the thermodynamic limit) ground states under the open boundary condition (OBC). The model undergoes the topological transition at $\theta = -\frac{\pi}{4}$ to a gapped dimerized phase with doubly degenerate (in the thermodynamic limit) ground states under the PBC and a single ground state under the OBC for an even number of spins. At the critical point $\theta = -\frac{\pi}{4}$, the model is integrable via the Bethe ansatz, and is known as the Takhtajan-Babudjan model [24,25]. The critical model is gapless and the spin-spin correlation function exhibits power-law behavior. The low-energy physics is described by the $SU(2)_2$ Wess-Zumino-Witten (WZW) model [18], with a central charge $c = 3/2$.

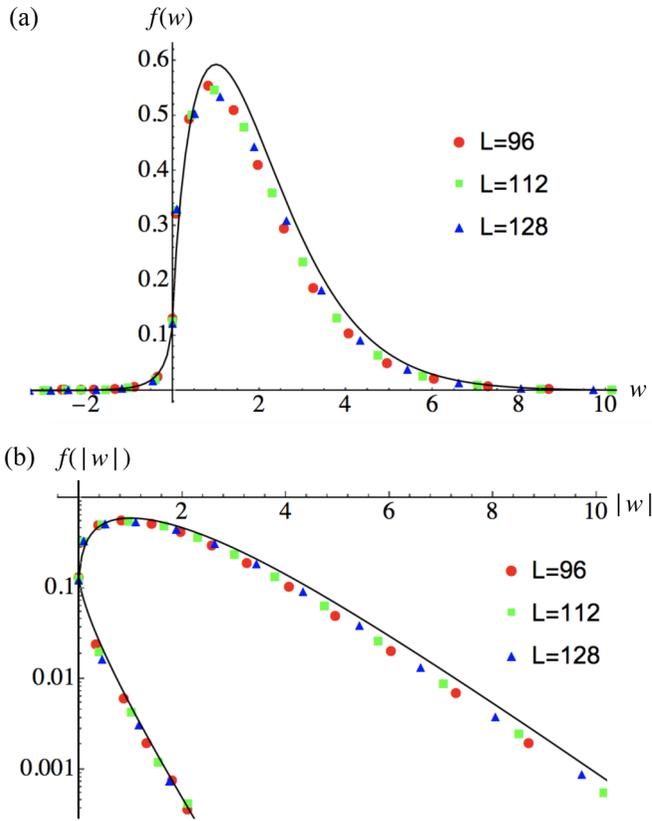

FIG. 3. (a) Numerical results for scaling function $f(w)$ of the bilinear-biquadratic spin-1 chain, Eq. (7), for three system sizes. The black solid line is Eq. (6). Positive (negative) $w$ represents the Haldane (dimerized) phase. (b) Same data in a log $f$ vs $|w|$ plot. The upper (lower) branch corresponds to the Haldane (dimerized) phase. Notice that the scaling function $f(w)$ decays as $e^{-|w|}$ in the Haldane phase and as $e^{-2|w|}$ in the dimerized phase for $|w| \gg 1$.

We evaluate the ground-state energy and the correlation length of the model with the help of the variational MPS approach and the VUMPS algorithm. Details of the method and the scaling function evaluation are described in the next section. Our results are presented in Fig. 3. Even for our largest systems there is still a slow size dependence of the scaling function. It very much looks like it tends to converge to the limiting form, given by noninteracting fermions, Eq. (6).

This result is not entirely surprising. Based on the Affleck-Haldane realization [18–20] that the critical point is described by $c = 3/2$ CFT, Tsvelik argued [21] that the vicinity of the transition may be described by three species of massive Majorana spinors, $\chi^a$, where $a = 1, 2,$ and 3. This statement is based on the analysis of relevant perturbations around the $SU(2)_2$ conformal point. For level-$k$ WZW theory, the primary fields are classified by their spin representation $j$ and have conformal dimensions $2j(j+1)/(2+k)$. For $k = 2$, the spin-1/2 field has the dimension 3/8, while the dimension of the spin-1 perturbation is 1. The former is nonlocal in Majorana fields and odd upon translation by one lattice site. As a result, it cannot be present in translationally invariant models. The spin-1 perturbation, on the other hand, is even under translations, local and quadratic in Majorana's, $\bar{\chi}^a \chi^a$. There is also a composite marginal (dimension 2) operator allowed by the symmetries of the form $J_\mu^a J_\mu^a$, where the chiral currents are $J_\mu^a = i\varepsilon^{abc}\bar{\chi}^b \gamma_\mu \chi^c$ and $\gamma_0 = \sigma_x$ and $\gamma_1 = i\sigma_y$. As a result, the low-energy Lagrangian close to the transition acquires the form

$$\mathcal{L} = i\bar{\chi}^a \gamma_\mu \partial_\mu \chi^a - m\bar{\chi}^a \chi^a - \lambda J_\mu^a J_\mu^a, \quad (8)$$

where the mass $m \propto \Delta\theta = \theta + \pi/4$ and $\lambda$ is a marginal coupling. As shown in Ref. [22], the role of the marginal four-fermion term is to renormalize the excitation gap $\Delta$ as

$$\Delta = m(1 + \lambda a \log m), \quad (9)$$

where $a$ is of the order of the lattice spacing. We show in the next section that the inverse correlation length, $1/\xi \propto \Delta$, may be indeed reasonably well fit with with this expression.

After the renormalization, Eq. (9), the low-energy spectrum of the model is given by the three species of Majoranas (with the renormalized mass). Remarkably this statement goes beyond the bulk of the spectrum, but also encompasses the finite-size effects, including the energies of the edge states. This illustrates a remarkable universality of $1 + 1$ topological transitions. This universality is not limited to the transition point, but extends away from it as long as the correlation length is large. The peculiarities of individual models are packed into a number of Majoranas and a specific (nonuniversal) dependence of their correlation length (inverse excitation gap) on the parameters.

## III. CHAINS WITH ODD NUMBER OF SPINS

The scaling function in odd-size chains appears to be significantly different from that in even-sized ones (Fig. 4). Below we explain that this difference may be accounted for in a straightforward way, which is however distinct on the two sides of the topological transition. The average bulk energy per spin, $\bar{\epsilon}(\xi)$, is, of course, the same for even and odd chains. One should be more careful, though, regarding the boundary term, $b(\xi)$. In the Haldane phase, where each spin-1 may be thought of as two spin-1/2 with bonding between two spin-1/2 at neighboring sites [Fig. 2(a)], the boundary term is insensitive to the parity of the chain. In the dimerized phase [Fig. 2(b)], however, there is clearly a difference depending on whether all spin-1's can be dimerized (even) or one is left "free" (odd) (Fig. 5).

An unpaired spin in the dimerized phase of the odd chain is actually a magnon excitation with the dispersion relation

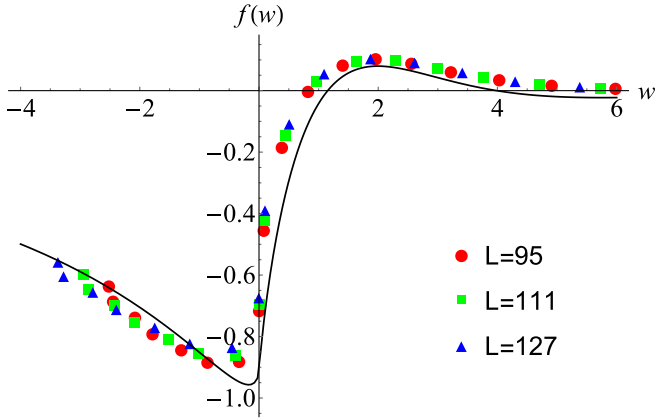

FIG. 4. Scaling function $f(w)$ of open odd chains with different system sizes. The black solid line is the analytical result. Its $w < 0$ and $w > 0$ parts are given by Eqs. (10) and (12), respectively.

$\varepsilon_k = \sqrt{\Delta^2 + k^2}$. For a finite-odd-size open chain the momentum quantization is given by the condition $e^{2ikL} = -1$, leading to $k_n = (n + 1/2)\pi/L$. The ground state corresponds thus to $k_0 = \frac{\pi}{2L}$, resulting in the ground-state energies $E_{\text{odd}} = E_{\text{even}} + \varepsilon_{k_0}$, which leads to the scaling function for $w < 0$:

$$f_{\text{odd}}(w) - f_{\text{even}}(w) = \lim_{L \to \infty} \frac{L}{c}\left[\Delta - \sqrt{\Delta^2 + \left(\frac{\pi}{2L}\right)^2}\right]$$

$$= \frac{2|w|}{3} - \sqrt{\left(\frac{2w}{3}\right)^2 + \left(\frac{\pi}{3}\right)^2}, \quad (10)$$

where we employ $c = \frac{3}{2}$ and $w = L\Delta$. This gives for the critical point $f_{\text{odd}}(0) = -\frac{7\pi}{24}$, which agrees with the spin-1 Kondo model [26] and with our numerical result within 3% error [27]. Equation (10) fairly well agrees with the variational matrix product data, as can be seen on the $w < 0$ side of Fig. 4.

On the Haldane side of the topological transition, $w > 0$, the difference between odd and even cases may be understood in terms of the edge spin-1/2 degrees of freedom [28,29]. For a finite-size chain there is a residual exponentially weak interaction between the edge spins given by [28,30]

$$E_{\text{int}} = (-1)^L \frac{a}{L} e^{-L/2\xi} \vec{s}_1 \cdot \vec{s}_2, \quad (11)$$

where $a$ is a numerical constant to be determined below and $1/L = v_F/L$ is the energy scale of the interactions. The $(-1)^L$ factor tells that the ground state alternates between the singlet and the degenerate triplet in even and odd cases, correspondingly. From here one finds for the difference of the ground-state energies $E_{\text{int}}^{\text{odd}} - E_{\text{int}}^{\text{even}} = \frac{a}{L} e^{-L/2\xi}$. This leads to the scaling function difference for $w > 0$,

$$f_{\text{odd}}(w) - f_{\text{even}}(w) = -\frac{\pi}{3} e^{-w/2}, \quad (12)$$

where we fixed $a = \pi/2$ to have the correct value $f_{\text{odd}}(0) = -\frac{7\pi}{24}$, established above [recall that $f_{\text{even}}(0) = \frac{\pi}{24}$]. This result is plotted in Fig. 4 on the $w > 0$ side. It agrees well with the variational matrix product simulations.

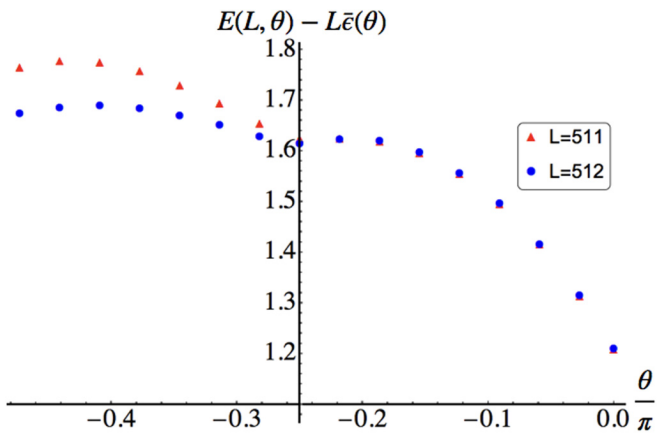

FIG. 5. The difference $E(L, \theta)/L - \bar{\epsilon}(\theta)$ vs $\theta$ for even and odd sizes $L$. The two coincide in the Haldane phase, but are distinct on the dimerized side. This illustrates that the boundary term $b(\xi)$ is sensitive to the chain parity in the dimerized phase.

We conclude that Eqs. (10) and (12) [where $f_{\text{even}}(w)$ is given by Eq. (6)] fully account for the difference between odd and even scaling functions on the dimerized and Haldane sides of the topological transition, correspondingly.

## IV. ALGORITHMS

### A. Variational MPS approach

Here we provide a brief recap of the variational MPS approach [13–15]. Consider a one-dimensional chain of $L$ sites and $d$-dimensional local state space $|\sigma_i\rangle$ on site $i$. For interacting systems, the Hilbert space of the chain grows exponentially with the number of sites. A generic pure many-body state is

$$|\psi\rangle = \sum_{\sigma_1, \cdots, \sigma_L} c_{\sigma_1, \ldots, \sigma_L} |\sigma_1, \ldots, \sigma_L\rangle, \quad (13)$$

with $d^L$ coefficients $c_{\sigma_1, \ldots, \sigma_L}$. One can find a more local representation of the state by using singular value decomposition (SVD); for any arbitrary rectangular matrix $M$ there exists SVD: $M = USV^\dagger$. Suppose $M$ is of dimension $m \times n$, then $U$ is of dimension $m \times \min(m, n)$ and is left normalized, i.e., $U^\dagger U = I$; $V$ is of dimension $n \times \min(m, n)$ and is right normalized, i.e., $VV^\dagger = I$; and $S$ is a diagonal matrix of dimension $\min(m, n)$ with non-negative entries $s_a$, called singular values.

By applying successive SVDs to the array of coefficients, the quantum state in Eq. (13) can be represented as a product of local tensors, or the so-called *matrix product state*:

$$|\psi\rangle = \sum_{\sigma_1, \cdots, \sigma_L} M^{\sigma_1} M^{\sigma_2} \cdots M^{\sigma_{L-1}} M^{\sigma_L} |\sigma_1, \ldots, \sigma_L\rangle. \quad (14)$$

The tensor $M^{\sigma_i}_{a_{i-1}, a_i}$ on site $i$ has three indices. Here $\sigma_i$ is a physical index, which corresponds to the dimension of the local state space $d$. While $a_{i-1}, a_i$ are two auxiliary indices. They count the left and right bonds through which the local state is connected to the left and right neighboring sites. The dimension of the bonds blows up exponentially with the distance to the edges: $\dim(\text{bond } a_i) = \min(d^i, d^{L-i})$, which means the decomposition itself does not reduce the complexity of calculation. In order to avoid exponential growth, it is demanded that the bond dimensions have a ceiling of $D$. The auxiliary space can be truncated due to the fact that the singular values of matrix $M^{\sigma_i}$ decay very fast. Therefore the exact SVD can be replaced by an approximate one: $M_{a_{i-1},a_i} \approx \sum_{a=1}^{D} U_{a_{i-1},a} S_{a,a} V^\dagger_{a,a_i}$. Note that the summation index $a$ runs over the largest $D$ singular values instead of $\min[\dim(\text{bond } a_{i-1}), \dim(\text{bond } a_i)]$. With this approximation, the bond dimension of the MPS representation is limited by $D$. The fast decay of singular values is guaranteed by the area law [31–35] for gapped systems; in critical systems, the decay is slower and the choice of $D$ depends on the system size. Also one should keep in mind that, in general, tensor $M^\sigma$ are different on each site.

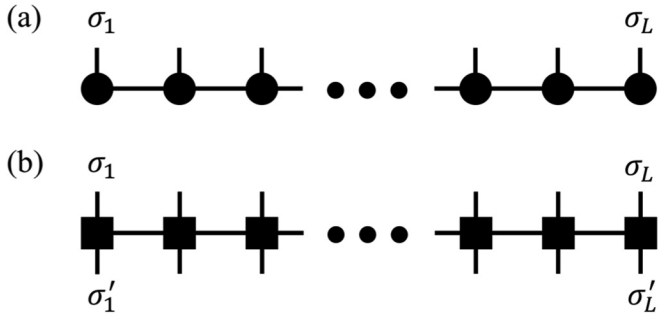

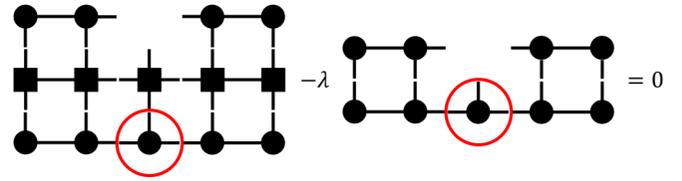

FIG. 7. Graphical representation of the eigenvalue problem for optimization of a single-site tensor. The unknown tensor is circled with red color. Usually two-site tensors are used in practical calculations.

FIG. 6. Graphical representation of (a) a general matrix product state and (b) a general matrix product operator. Solid circles and squares represent local tensors. Vertical bonds represent physical indices while horizontal bonds represent auxiliary indices.

In a similar way, an arbitrary operator $\hat{O}$ can be brought to a *matrix product operator* (MPO) form (Fig. 6):

$$\hat{O} = \sum_{\boldsymbol{\sigma},\boldsymbol{\sigma}'} W^{\sigma_1,\sigma_1'} W^{\sigma_2,\sigma_2'} \cdots W^{\sigma_{L-1},\sigma_{L-1}'} W^{\sigma_L,\sigma_L'} |\boldsymbol{\sigma}\rangle\langle\boldsymbol{\sigma}|, \quad (15)$$

where $|\boldsymbol{\sigma}\rangle = |\sigma_1, \ldots, \sigma_L\rangle$. The only difference is that the tensor $W^{\sigma_i,\sigma_i'}_{b_{i-1},b_i}$ on site $i$ has two physical indices and two auxiliary indices.

The search for the ground state of a Hamiltonian $\hat{\mathcal{H}}$ is equivalent to finding an optimal approximation of MPS $|\psi\rangle$ of dimension $D$ that minimizes the energy:

$$E = \frac{\langle\psi|\hat{\mathcal{H}}|\psi\rangle}{\langle\psi|\psi\rangle}. \quad (16)$$

An efficient algorithm to realize it is by doing variational search in the MPS space. To be more specific, we keep all but a tensor of small number of sites (usually one or two) constant, then take the extreme of $\langle\psi|\hat{\mathcal{H}}|\psi\rangle - \lambda\langle\psi|\psi\rangle$ with respect to the selected tensor, where $\lambda$ is a Lagrangian multiplier. This is equivalent to solving an eigenvalue problem whose eigenvalue $\lambda$ is the current ground-state energy and eigenvectors give the updated tensor (Fig. 7). When the updates are done iteratively through the entire chain from one end to the other end, it is said a sweep is completed. One continues doing sweeps along the chain until convergence of energy is reached.

Variational MPS calculations in this paper were performed using the ITensor Library [36].

### B. VUMPS algorithm

We now recap the VUMPS algorithm [16], which deals with systems in the thermodynamic limit $L \to \infty$. In this case, the ground-state approximation is constructed by a translation invariant uniform MPS, i.e., the same single MPS tensor $M^\sigma$ (or a unit cell of several tensors) on all sites:

$$|\psi\rangle = \sum_{\boldsymbol{\sigma}} \left(\cdots M^{\sigma_{i-1}} M^{\sigma_i} M^{\sigma_{i+1}} \cdots\right) |\boldsymbol{\sigma}\rangle. \quad (17)$$

By local gauge transformation, the above state can be brought into a left/right canonical representation with left/right normalized tensors $A/B$ that satisfy:

$$\sum_\sigma A^{\sigma\dagger} A^\sigma = \mathbb{1}, \quad \sum_\sigma A^\sigma \rho_A A^{\sigma\dagger} = \rho_A,$$

$$\sum_\sigma B^\sigma B^{\sigma\dagger} = \mathbb{1}, \quad \sum_\sigma B^{\sigma\dagger} \rho_B B^\sigma = \rho_B. \quad (18)$$

Here $\rho_A$ and $\rho_b$ are the reduced density matrices of the bipartited system.

With the help of left and right normalized tensors, we cast the state into a mixed canonical representation (Fig. 8):

$$|\psi\rangle = \sum_{\boldsymbol{\sigma}} \left(\cdots A^{\sigma_{i-1}} M_C^{\sigma_i} B^{\sigma_{i+1}} \cdots\right) |\boldsymbol{\sigma}\rangle, \quad (19)$$

where the center site tensor $M_C^\sigma$ is related to the left/right normalized tensor by a bond matrix $C$:

$$M_C^\sigma = A^\sigma C = C B^\sigma. \quad (20)$$

In fact, bond matrix $C$ relates the left and right normalized tensors $A$ and $B$ by a gauge transformation $A^\sigma = CB^\sigma C^{-1}$ and allows the arbitrary shift of the center site tensor on the chain. Furthermore, by applying the normalization condition and the fixed point relation in Eq. (18), one can verify that $\rho_A = CC^\dagger$ and $\rho_B = C^\dagger C$.

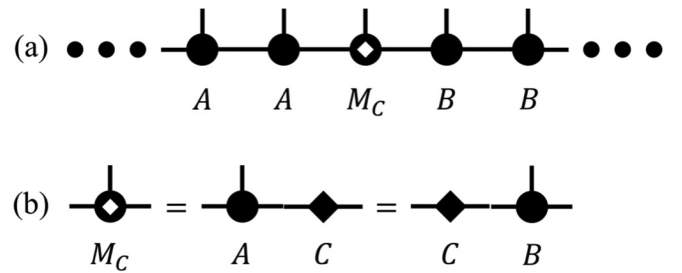

FIG. 8. (a) Graphical representation of the ground state in the mixed canonical representation. (b) Relationship between the center matrix $M_C$ and the bond matrix $C$.

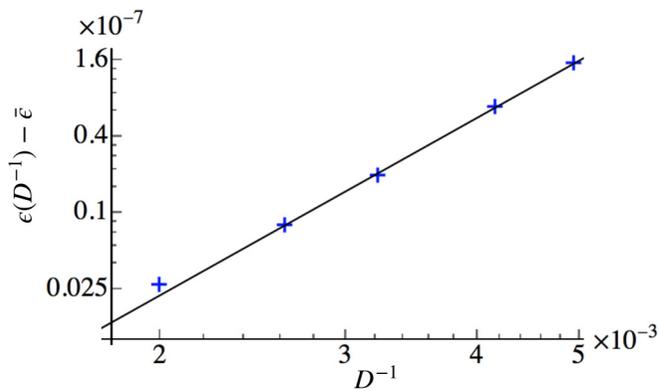

FIG. 9. Log-log plot of the variational ground-state energy of uniform MPS as a function of inverse bond dimension. The blue "+" symbols represent the numerical data (at $\theta = -\pi/4 + 0.3$). The plot follows the power-law fit $\epsilon(D^{-1}) = \bar{\epsilon} + a(D^{-1})^b$, where $a$ and $b$ are fitting parameters.

In order to find the ground state in the thermodynamic limit, we again apply the Hamiltonian to the uniform MPS we constructed in Eq. (19) and solve the effective eigenvalue problem. But instead of sweeping through the entire chain (which destroys the translational symmetry of the state), we only solve for the center site tensor $M_C^\sigma$ and the bond matrix $C$. We then compute the left and right normalized tensors by Eq. (20) and update the state globally. Convergence is considered to be reached when the tensor $M_C^\sigma$ no longer changes.

### C. Details of the simulations

The ground-state energy of the finite-size system is calculated by the regular variational MPS approach with truncation error at order $10^{-12}$. To obtain the average bulk energy density $\bar{\epsilon}$, we consider the ground-state energy of uniform MPS at different bond dimensions $D$. By plotting the energy density $\epsilon$ as a function of inverse bond dimension and fitting the relationship using a power law, $\bar{\epsilon}$ is extrapolated by letting bond dimension $D \to \infty$ (see Fig. 9). The average ground-state energy per spin, $\bar{\epsilon}$, is plotted as a function of the tuning parameter $\theta/\pi$ in Fig. 10 (a similar result was obtained in Ref. [37]). At criticality, $\theta/\pi = -1/4$, the ground-state energy is known [24,25] from the Bethe ansatz to be $\bar{\epsilon} = -2\sqrt{2} = -2.828\,427$; our numerical result is $-2.828\,426$.

The boundary term $b$ is found by subtracting the total bulk energy from the ground-state energy for some large system size. Figure 11 shows that $E(L, \theta/\pi) - L\bar{\epsilon}(\theta/\pi)$ gets saturated as the system size increases. The limit size used in this paper is $L = 512$.

The correlation length can be calculated through the eigenvalues of the transfer matrix of the uniform MPS state. The transfer matrix is defined as

$$T = \sum_\sigma \bar{M}^\sigma \otimes M^\sigma, \quad (21)$$

where $M^\sigma$ is the repeated tensor on each site in Eq. (17). One can prove that the eigenvalue of the transfer matrix is bounded by 1 given that the wave function is normalized [15]. Suppose that the eigenvalues of $T$ are sorted in descending order $\lambda_1 > \lambda_2 \geqslant \lambda_3 \geqslant \cdots$, with $\lambda_1 = 1$ being nondegenerate. Then the correlation function between two operators $O$ with distance $j - i$ is

$$\langle \psi | \hat{O}^{[i]} \hat{O}^{[j]} | \psi \rangle = \cdots T^{[i-1]} T_O^{[i]} T^{[i+1]} \cdots T^{[j-1]} T_O^{[j]} T^{[j+1]} \cdots$$
$$= \sum_k \langle 1 | T_O^{[i]} | k \rangle \lambda_k^{j-i-1} \langle k | T_O^{[j]} | 1 \rangle. \quad (22)$$

Here $|k\rangle$ and $\langle k|$ are the right and left eigenvectors of transfer matrix $T$ which corresponds to eigenvalue $\lambda_k$. Thus the correlator is a superposition of exponentials with the decay lengths $\xi_k = -1/\ln \lambda_k$. And the MPS two-point correlation

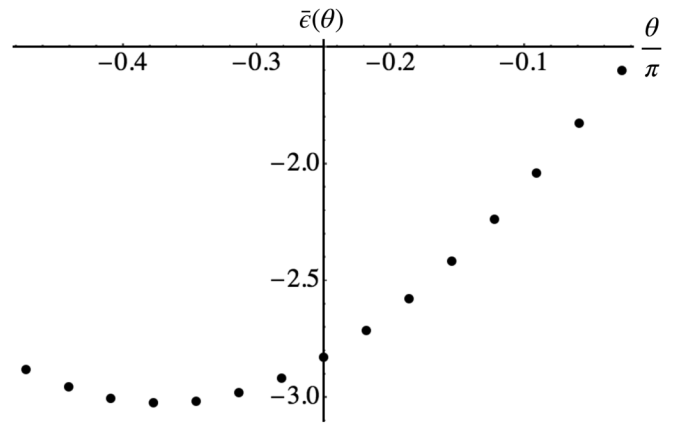

FIG. 10. The bulk energy per spin, $\bar{\epsilon}$, vs parameter $\theta$. The critical point is at $\theta/\pi = -1/4$. The left (right) side of the critical point is the dimerized (Haldane) phase.

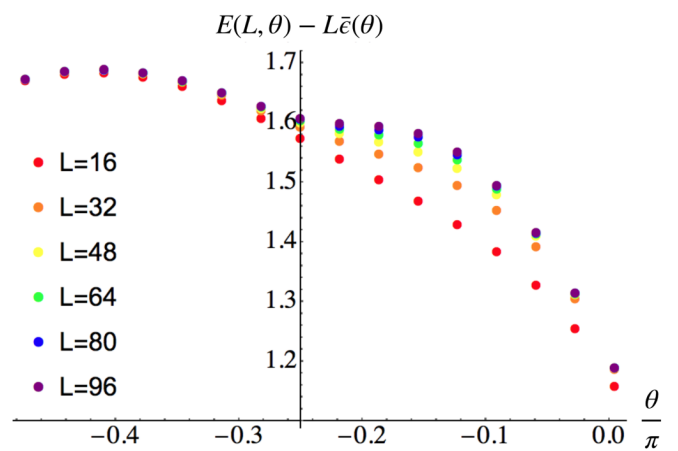

FIG. 11. The difference $E(L, \theta) - L\bar{\epsilon}(\theta)$ vs $\theta$ plotted for different system sizes $L$. As the system size increases the subleading term $cf(L/\xi)/L$ is negligible. The difference thus saturates to the boundary term $b(\theta)$ [cf. Eq. (3)].

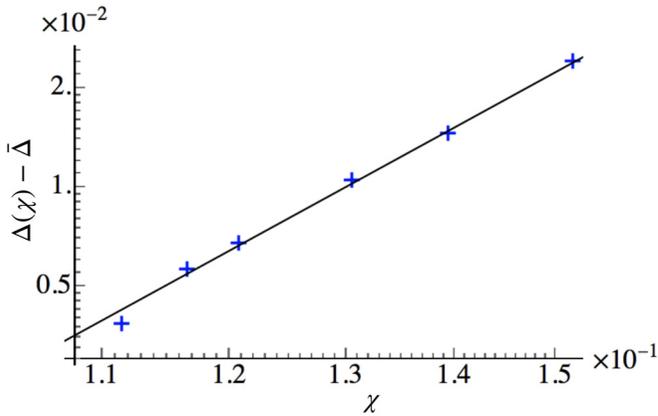

FIG. 12. Log-log plot of the inverse of the correlation length $\Delta$ as a function of the inverse of $\log(D)$. The blue "+" symbols represent the numerical data (at $\theta = -\pi/4 + 0.3$). The plot follows the power law fit $\Delta(\chi) = \bar{\Delta} + a\chi^b$, where $a$ and $b$ are fitting parameters.

function have the following generic form:

$$\frac{\langle\psi|\hat{O}^{[i]}\hat{O}^{[j]}|\psi\rangle}{\langle\psi|\psi\rangle} = c_1 + \sum_{k=2}^{D} c_k e^{-r/\xi_k}, \quad (23)$$

where $r = |j - i - 1|$ and $c_k = \langle 1|T_O^{[i]}|k\rangle\langle k|T_O^{[j]}|1\rangle$. The correlation length is defined by the largest decay length $\xi = -1/\ln\lambda_2$. As the bond dimension $D$ of tensor $M^\sigma$ increases, the correlation length is saturated. To be more specific, the inverse of the correlation length is related to the inverse of $\log D$ by a power law [38]. If we denote $\Delta = \xi^{-1}$ and $\chi = (\log D)^{-1}$, then $\Delta(\chi) = \bar{\Delta} + a\chi^b$ as is shown in Fig. 12.

The calculated inverse correlation length $\xi^{-1}$ vs $\Delta\theta$ is shown in Fig. 13. In Fig. 14 we fit $(\xi\Delta\theta)^{-1}$ with $a_1 + a_2\log\Delta\theta$, which is the expected dependence due to the renormalization by the marginal composite four-Majorana operator, Eq. (9). The fit is rather satisfactory, though we do not know how to independently verify the fitting parameters $a_1$ and $a_2$ on the two sides of the transition.

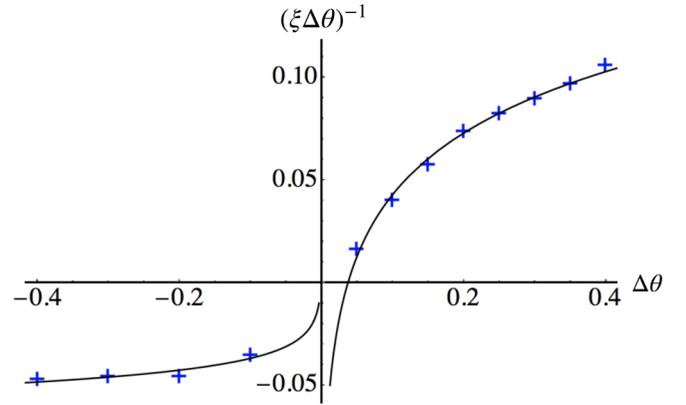

FIG. 14. $(\xi\Delta\theta)^{-1}$ vs deviation from the critical point $\Delta\theta = \theta + \pi/4$. Here $\Delta\theta > 0$ ($<0$) represents the Haldane (dimerized) phase. The blue "+" symbols represent the numerical data. Data are fitted with $a_1 + a_2 \log \Delta\theta$. As mentioned in the main text, the fitting parameters $a_1$ and $a_2$ are different on two sides of the transition: $|a_1| = 0.056$ and $|a_2| = 0.0083$ on the left (dimerized phase) and $a_1 = 0.14$ and $a_2 = 0.043$ on the right (Haldane phase).

Finally we show in Fig. 15 the convergence of the scaling function, Eq. (3), towards the free Majorana result, Eq. (6), for small system sizes [the larger sizes are shown in Fig. 3(a)].

## V. CONCLUSION AND OUTLOOK

In this paper, we study the finite-size scaling function across the topological phase transition in bilinear-biquadratic spin-1 chain. The conformal transition point is known [21] to be described by three copies of Majorana fermions. We have performed high-precision variational matrix product calculations and found that this description still holds away from the transition in the scaling regime, where the correlation length is much larger than the microscopic scales. Moreover, it correctly describes the finite-size effects, in particular, the universal finite-size scaling function [9]. We have also observed that there is a marked difference between even- and odd-size chains. This difference may be fully account for by simple considerations, incorporating the essential physics of the dimerized and the topological phases. In particular, in the topological Haldane phase it may be entirely attributed to the residual interactions of the edge spin-1/2 excitations.

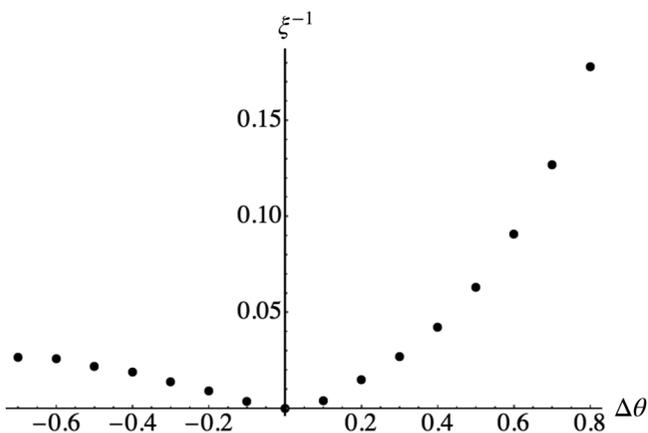

FIG. 13. Inverse correlation length $\xi^{-1}$ vs deviation from the critical point $\Delta\theta = \theta + \pi/4$. Here $\Delta\theta > 0$ ($<0$) represents the Haldane (dimerized) phase.

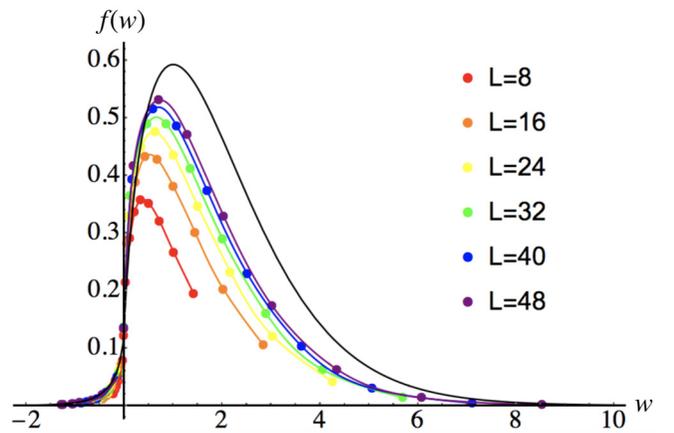

FIG. 15. Scaling function $f(w)$ is plotted, where $w = L/\xi$. Points of different colors represent calculations for different system sizes. The solid lines are guides for the eye. The black solid line is the analytical result, Eq. (6).

In finite-size systems the fourfold degeneracy of the Haldane phase for the OBC is lifted with the splitting $\sim\exp(-L/2\xi)$. In even chains the ground state is a unique, singlet state of the two edge spin-1/2 excitations, while in odd chains it is the triple degenerate triplet. In the dimerized phase the ground state is again unique (spin zero) in even chains and is again the triplet in odd chains. Thus, for the finite-size case the exact degeneracy of the ground state is not changed at the transition, as it should be. Away from the transition, extra states may become exponentially close to the ground state. The scaling function probes the restoration of the thermodynamic limit degeneracy away from the transition.

For the PBC, the Haldane phase is gapped with a unique ground state. In the dimerized phase the ground state, while unique for the finite-size case (and an even number of spins), becomes twofold degenerate in the thermodynamic limit. This may indicate a nonsymmetric scaling function under the PBC. On the other hand, the critical theory of the three Majorana fermions exhibits a perfectly symmetric scaling function for both even- and odd-sized cases [9]. It would be interesting to simulate a large enough PBC model to resolve this issue.

It would be also instructive to check the scaling function in other interacting systems, where the low-energy field theory may not be described by free fermions. Examples include the spin-1 XXZ chain with the uniaxial single-ion-type anisotropy [39–41] and the bond-alternating spin-1/2 Heisenberg chain [42].

## ACKNOWLEDGMENTS

We are grateful to I. Affleck, A. Chubukov, T. Gülden, M. Stoudenmire, A. Tsvelik, and V. Zauner-Stauber for useful comments and discussions. This work was supported by NSF Grant No. DMR-1608238.

## APPENDIX: DETAILS OF THE VARIATIONAL MPS APPROACH

The MPO representation of Hamiltonian (7) is shown below:

$$\hat{\mathcal{H}} = \hat{W}^{[1]}\hat{W}^{[2]}\cdots\hat{W}^{[L]}, \quad (A1)$$

where matrix $\hat{W}^{[i]}$, $i \in [2, L-1]$, has zero entries except the first column $\hat{W}^{[i]}_{All,1}$ and the last row $\hat{W}^{[i]}_{14,All}$:

$$\hat{W}^{[i]}_{All,1} = [\,\hat{I} \quad \hat{S}^- \quad \hat{S}^+ \quad \hat{S}^z \quad \hat{S}^{-2} \quad \hat{S}^-\hat{S}^+ \quad \hat{S}^-\hat{S}^z$$
$$\hat{S}^+\hat{S}^- \quad \hat{S}^{+2} \quad \hat{S}^+\hat{S}^z \quad \hat{S}^z\hat{S}^- \quad \hat{S}^z\hat{S}^+ \quad \hat{S}^{z2} \quad 0\,]^\dagger,$$

$$\hat{W}^{[i]}_{14,All} = \left[\,0 \quad \frac{\cos\theta}{2}\hat{S}^+ \quad \frac{\cos\theta}{2}\hat{S}^- \quad \cos\theta\hat{S}^z \quad \frac{\sin\theta}{4}\hat{S}^{+2}\right.$$
$$\frac{\sin\theta}{4}\hat{S}^+\hat{S}^- \quad \frac{\sin\theta}{2}\hat{S}^+\hat{S}^z \quad \frac{\sin\theta}{4}\hat{S}^-\hat{S}^+ \quad \frac{\sin\theta}{4}\hat{S}^{-2}$$
$$\left.\frac{\sin\theta}{2}\hat{S}^-\hat{S}^z \quad \frac{\sin\theta}{2}\hat{S}^z\hat{S}^+ \quad \frac{\sin\theta}{2}\hat{S}^z\hat{S}^- \quad \sin\theta\hat{S}^{z2} \quad \hat{I}\,\right].$$

The matrices on two ends are as follows: $\hat{W}^{[1]} = \hat{W}^{[i]}_{All,1}$ is a pure row matrix and $\hat{W}^{[L]} = \hat{W}^{[i]}_{14,All}$ is a column matrix.